\documentclass[graybox]{svmult}

\usepackage[square,sort,comma,numbers]{natbib}
\usepackage{graphicx}        
\usepackage{float}
\usepackage{hyperref}
\usepackage{titlesec}
\usepackage{times}

\usepackage{makeidx}         
\usepackage{multicol}        
\usepackage[bottom]{footmisc}

\usepackage{newtxtext}       %
\usepackage[varvw]{newtxmath}       

\usepackage[margin=0.85in,bottom=1in,top=1in]{geometry}

\mathchardef\mhyphen="2D

\newcommand{\ltsima}{$\; \buildrel < \over \sim \;$}
\newcommand{\lsim}{\lower.5ex\hbox{\ltsima}}
\newcommand{\gtsima}{$\; \buildrel > \over \sim \;$}
\newcommand{\gsim}{\lower.5ex\hbox{\gtsima}}
\newcommand{\eg}{{\it e.g.,\ }}

\newcommand{\lcdm}{$\Lambda$CDM}

\def\muscle{{\scshape muscle}}

\newcommand{\origami}{{\scshape origami}}
\newcommand{\arepo}{{\scshape Arepo}}

\def\gtrsim{\mathrel{\hbox{\rlap{\hbox{\lower4pt\hbox{$\sim$}}}\hbox{$>$}}}}

\def\lesssim{\mathrel{\hbox{\rlap{\hbox{\lower4pt\hbox{$\sim$}}}\hbox{$<$}}}}

\def\eg{e.g.\ }

\definecolor{ForestGreen}{rgb}{0.3,0.7,0.3}

\title{Boundaries of determinism and chaos in the cosmos}

\begin{document}

\title*{Boundaries of chaos and determinism in the cosmos}
\author{Mark Neyrinck, Shy Genel and Jens St\"ucker}
\institute{Mark Neyrinck~(\email{Mark.Neyrinck@gmail.com})\at Ikerbasque, Basque Foundation for Science\\
Department of Physics, University of the Basque Country UPV/EHU, Bilbao, Spain\\
Donostia International Physics Center
\and Shy Genel \at Center for Computational Astrophysics, Flatiron Institute\newline
Columbia Astrophysics Laboratory, Columbia University
\and Jens St\"ucker \at Donostia International Physics Center}
%
%
\maketitle

\abstract{According to the standard model of cosmology, the arrangement of matter in the cosmos on scales much larger than galaxies is entirely specified by the initial conditions laid down during inflation. But zooming in by dozens of orders of magnitude to microscopic (and human?) scales, quantum randomness reigns, independent of the initial conditions. Where is the boundary of determinism, and how does that interplay with chaos? Here, we make a first attempt at answering this question in an astronomical context, including currently understood processes. The boundary is a function, at least, of length scale, position, and matter type (dark matter being more simply predictable). In intergalactic voids, the primordial pattern of density fluctuations is largely preserved. But we argue that within galaxies, the conditions are at minimum chaotic, and may even be influenced by non-primordial information, or randomness independent of the initial conditions. Randomness could be supplied by events such as supernovae and jets from active galactic nuclei (AGN) and other accretion disks, which, with the help of chaotic dynamics, could broadcast any possible microscopic randomness to larger scales, eventually throughout a galaxy. This may be generated or amplified by a recently investigated process called spontaneous stochasticity, or effective randomness in turbulent systems arising from arbitrarily small perturbations.}

\section{Introduction}
To what degree is the current universe determined from the primordial universe? Given a perfect map of the primordial observable universe shortly after the Big Bang, what is predictable from it? What was encoded there, and to what accuracy? This paper? The COVID-19 pandemic? Life on Earth? Earth itself? The Sun? The Milky Way galaxy?

A consensus on an observationally successful concordance \lcdm\ cosmological model (cold dark matter with a cosmological constant) has emerged rather recently, enabling us to say some things with confidence about the philosophical question of the degree of determinism in the cosmos. Determinism is a substantial topic in philosophy, associated with topics such as the existence of free will. But this paper does not have clear implications for free will, particularly since philosophers disagree about whether determinism and free will are compatible ($\sim 59\%$ of philosophers think they are compatible, according to a 2014 survey \cite{BourgetChalmers2014}).

In general, determinism may be partial, i.e.\ the final conditions can depend both on the initial conditions, and on subsequently added, non-primordial information that has no dependence on the initial conditions. In a broad class of Big Bang cosmological models, including \lcdm, some things at late times are determined almost perfectly from the initial conditions (ICs), and some not. `The ICs' are important enough to explicitly define here; they are what would constitute input into a cosmological large-scale structure simulation: a 3D map of the primordial density and velocity fields. In the standard inflationary model, these are primordial random quantum fluctuations, a Gaussian random field to a good approximation. These `inflated' to macroscopic size during inflation, in the first fraction of a second. The initial conditions are `primordial randomness' -- the big dice roll that determined much of what followed. Gravity took this initial density map as a blueprint for where to form structures.

But on non-astronomical scales, e.g.\ on Earth, it seems that `non-primordial randomness' dominates. This is information, or stochasticity, that has no direct dependence on the ICs. In this paper, we assume that this information is truly random, not determined by hidden deterministic degrees of freedom. The regime dominated by non-primordial randomness encompasses scales of the weather on Earth and smaller, but clarifying the boundary in scale and location between primordial and non-primordial randomness is not trivial, and is the aim of this paper.

Quantum randomness is obviously present on microscopic scales, but it influences larger scales as well. Prime examples of random quantum processes are the emission of a photon by an excited atom, or the radioactive decay of an atomic nucleus. These decays occur in elements such as uranium within the Earth, and also in faraway systems, producing cosmic rays that hit the Earth. Encounters with such radioactivity cause mutations in organisms, affecting biological evolution \cite[e.g.][]{AtriMelott2014} which ended up being consequential for the Earth's surface and its climate. If a mutation is from a cosmic ray, where, when, and whether it arrives are sensitive to random radioactive decays. Random cosmic ray strikes could even seed clouds \cite{SvensmarkEtal2017}, thus possibly serving as `butterflies' in the chaotic weather. The formation of life may also have arisen from random, low-probability events that were not prescribed in the initial conditions.

Going to larger scales, things are not so obviously random. The regular (on any human timescale) motions of the Solar System underlie the etymology of the orderly `cosmos.' But to what degree was the Solar System encoded in the ICs? Was it specified in detail, or is it merely a draw from the distribution of possible star systems that could occur in the ensemble of possible Milky Ways? How wide is the possible distribution of Solar Systems? Of Milky Ways? The more non-primordial randomness there is in the cosmos, the wider these distributions. While the position of possible Milky Ways was likely determined with high accuracy by the initial conditions, its internal structure and properties may have a wide distribution.

As we discuss below, we expect that the width of the distribution of fates goes up with the amount of crossings that a mass element (particle) undergoes. For very massive particles, at least the mass of large galaxies, there is little room for randomness to affect their position and character. As the size of the element decreases, the distribution width becomes larger, and possibly position-dependent. Particles much smaller than galaxies will have orbited many times if they are within a galaxy. These dynamical crossing times are crucial to the excitation of chaos, which could amplify small amounts of randomness to large scales.

Deterministic chaos is different from randomness, but they can work together; chaos could help a small random fluctuation expand to large scale and amplitude. Recent research has found that turbulent fluids can even produce effective randomness on macroscopic scales from arbitrarily small initial fluctuations; this is called spontaneous stochasticity. This is an extreme (but not necessarily rare) form of chaos, in which the magnitude of a difference in the final conditions remains large, for any tiny, nonzero perturbation to initial conditions.

This paper is structured as follows. In \S \ref{sec:largescales}, we discuss where we expect dynamics to be deterministic: large scales, and single-stream voids. In \S \ref{sec:dmosim}, we show results from a set of dark-matter-only simulations in which we explicitly find the degree of chaos in different regimes. \S \ref{sec:galaxies} discusses some reasons to expect large amounts of chaos, and perhaps even randomness, to occur in galaxies, including \S \ref{sec:ss}, which elaborates on spontaneous stochasticity. In \S \ref{sec:outflows}, we discuss further cosmological simulations with hydrodynamics, star-formation and feedback processes that try to measure whether galactic outflows might pollute voids with randomness. In \S \ref{sec:dynamicalinfo}, we speculate on how non-primordial randomness may come to dominate the cosmic information budget, and why this might be of cosmological interest. Finally, in \S \ref{sec:summary} we summarise the paper and conclude.

\section{Large scales, and voids: cosmos, not chaos}
\label{sec:largescales}
Smoothing over the universe on a sufficiently large length scale (or mass scale), the density field in the Universe is easy to predict. Densities in large pixels (many tens of Mpc) at the present epoch are well-described as being an increasing global constant (the growth factor) times the initial pixel densities at the same locations in comoving coordinates, i.e.\ following the mean expansion of the Universe. This prescription is called Eulerian linear perturbation theory.

But zooming in, the scatter in this relationship grows, until pixel densities become essentially uncorrelated with their initial values. One way to describe this is with the `propagator' \cite{CrocceScoccimarro2006} a (Fourier-space) cross-correlation between initial and final pixels. It is 1 on large scales, 0 on small scales, and transitions at a scale of $r_{\rm nonlinear}\approx 10$ Mpc/$h$ at the current epoch.

Increasing resolution to $\sim$1 Mpc$/h$ (about the Milky Way-Andromeda distance) reveals the cosmic web \cite[e.g.][]{VDWetal2016,LibeskindEtal2018}. On this length scale, there is essentially no correlation between initial and final comoving pixel densities. But it is more meaningful to talk about predictability in a Lagrangian sense, looking at where mass moves between the initial and final conditions. There are simple Lagrangian prescriptions that give particle positions in voids, and the boundaries of voids, to an accuracy surprising if one is used to thinking of structure formation as hopelessly black-box and nonlinear.

Even the first-order Zel'dovich Approximation (\cite{Zeldovich1970}, ZA, linear perturbation theory in particle displacements) gives the final structure rather accurately. The ZA is used in cosmological simulations to generate the IC files from the initial density map. More accuracy is possible, at the expense of a bit more computation (still much less than a full $N$-body simulation, considered the ground-truth prediction); see \cite{Monaco2016} for a review of these methods. The cross-correlation between ICs and final conditions in a Lagrangian approach is generally good, with a resolution nearly a factor of 10 better than in a linear Eulerian approach \cite{Neyrinck2016}.

\begin{figure}
\centering
    \includegraphics[width=0.8\columnwidth]{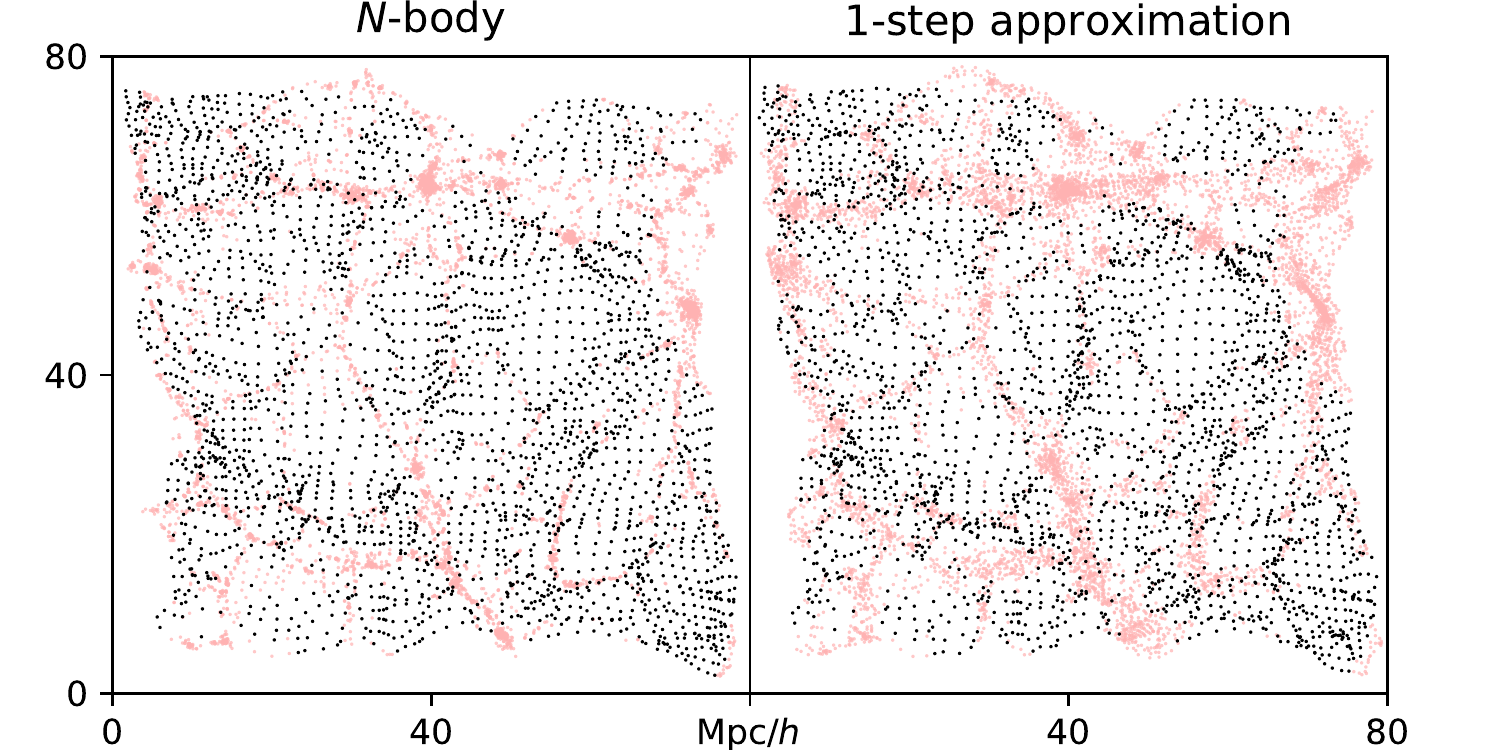}
\caption{\small A patch of 100$^2$ dark-matter particles from a set of 256$^3$ particles advanced in 3D from a set of initial conditions with an $N$-body code, and with a recent approximation, {\scshape muscle}. Fine details in the patchwork of void particles (black) are preserved in the final conditions, and reproduced in the approximation. It is so predictable in these regions because the patchwork has not folded up at all; the patches evolve nearly as independent `separate universes.' This implies a high amount of determinism for these particles, since the approximation responds only slightly (weaker than linearly) to a change in the initial conditions. In pink regions, the patchwork quilt has folded up to form more complicated collapsed structures that this prescription does not predict accurately, i.e.\ galaxies and filaments. The centers of these pink structures are still rather accurately predicted, though.}
  \label{fig:muscle_voids}
\end{figure}

Fig.\ \ref{fig:muscle_voids} shows a visual example of a recent Lagrangian method, \muscle, which is a one-step `multiscale spherical-collapse' prescription \cite{Neyrinck2016,TosoneEtal2021}. The figure shows a 2D projection of a 100$^2$ patch of dark-matter particles from a gravity-only $N$-body simulation\footnote{Of box size 200 Mpc/$h$, and 256$^3$ particles, run with the {\scshape gadget2} \cite{Gadget2} code assuming concordance cosmological parameters $(\Omega_\Lambda=0.7,\Omega_M=0.3,h=0.7)$.}, along with the \muscle\ prediction of particle locations from the same ICs.

\subsection{Length scale of predictability}
It is useful to define a `length scale of predictability' function $s_p(m,e,y,c)$ for a particle, where the first three arguments are a guess about a simple set of properties this depends on. It gives a particle's positional uncertainty (low $s_p$ means high accuracy) at some epoch. It could generally be a function of particle mass $m$; particle `environment' $e$ (the entire set of  position-dependent quantities such as final cosmic-web type relevant to determine $s_p$); the type of particle $y$ (e.g.\ dark or baryonic matter); and $c$, the amount of computation (in floating-point operations, say) allowed to do to reach a prediction. To be more precise, $s_p$ should really be the minimum uncertainly over all algorithms with an amount of computation $c$, but below we just assume (probably wrongly) that \muscle~is the most accurate possible algorithm of its computation amount. Visually, \muscle's prediction is better for voids than non-voids. In terms of $s_p$, where $c_{\rm muscle}$ is the low amount (compared to an $N$-body simulation) of computation required to run the \muscle~algorithm, $s_p(m,{\rm void}, {\rm dark~matter}, c_{\rm muscle})$ looks lower (more accurate) than $s_p(m,{\rm non\mhyphen void}, {\rm dark~matter}, c_{\rm muscle})$.

Indeed, in non-voids (pink in Fig.\ \ref{fig:muscle_voids}), particle positions in \muscle~have a median error of $s_p(m,{\rm non\mhyphen void}, {\rm dark~matter}, \allowbreak c_{\rm muscle})= 1.2\ {\rm Mpc}/h$, while in the voids (black), fine details of the structure often match and $s_p(m,{\rm void}, {\rm dark~matter}, c_{\rm muscle})\allowbreak = 0.5\ {\rm Mpc}/h$. The ZA is not much worse, giving $s_p = 1.3~{\rm Mpc}/h$ and $0.7~{\rm Mpc}/h$ for non-void and void particles, respectively. Void particles here are defined as not having crossed any other particles between the initial and final conditions, as identified in the $N$-body results with the {\scshape origami} definition \cite{FalckEtal2012}. 

The inaccuracy of the pink, non-void particle positions in the right panel of Fig.\ 1 suggests that more computation (or a better algorithm, or even the full $N$-body computation) is necessary to predict their positions accurately. We pause here to consider the expected behavior of $s_p$ as a function of allowed computation $c$. This computational cost is a type of complexity called `logical depth' \cite{Bennett1986,Bennett1990}. This complexity is complementary to the Kolomogorov information, which is the length of the shortest program capable of producing the output, without regard to the amount of computation required. If true randomness were added, this would be incompressible data that would constitute Kolmogorov information. But in the limit of arbitrarily high computation (logical depth), structures appear random, and logical depth becomes effectively indistinguishable from true randomness. In this definition, true randomness would appear as nonzero $s_p$ as the computation $c\to\infty$. In voids, $s_p$ seems to reach a low level even for low $c$, but it is less clear what happens in haloes, and for baryons.

\subsection{Simple physics in voids}
Why are void regions so simple? The key is the multistreaming definition of void regions here: the particles have not crossed any other particles. There are many other definitions of voids \cite[\eg][]{ColbergEtal2008}, but the \origami\ definition, related to a `sheet' view of dark-matter dynamics \cite{ShandarinEtal2012,AbelEtal2012}, is the most useful for our purposes. Conceptually, these are also the cells that remain uncollapsed in the adhesion model \cite[\eg][]{GurbatovEtal2012,HiddingEtal2012,NeyrinckEtal2018}.

Consider a patchwork of patches of slightly different density in the primordial Universe. If the patch does not collapse, to a good approximation it evolves as a `{\it separate universe},' with an expansion rate depending only on its density and the tidal field \cite{DaiEtal2015}. The broad validity of the separate-universe approximation explains why the dynamics within voids are largely insulated from faraway happenings. External fields induce motions of the patches, but only slightly affect their expansion or contraction.

So, the patchwork remains intact, albeit distorting in a simple way, and requiring relatively little computation to predict. This is why voids act as `cosmic magnifiers' of the primordial patterns inside them \cite{AragonCalvoSzalay2013,NeyrinckYang2013}. Not only are these single-stream regions predictable to a large degree with simple one-step prescriptions, but they dominate the Universe by volume, reaching well over 50\% \cite{AnguloWhite2010,StuckerEtal2018}.

Another way to see the lack of chaos in voids is to consider the (lack of) crossing time there. Ref.\ \cite{KandrupSmith1991} found that if nearby particles are perturbed a bit, the distance between them grows by a factor of $e$ on a timescale of order the particles' crossing or dynamical time in a virialized object. But voids are not just unvirialized, but have not even undergone a first collapse, or any matter crossing at all.

More quantitatively, consider the dynamics within a void, in a Lagrangian model. In the ZA, particle positions respond exactly linearly to changes in ICs, suggesting that chaos is not common there. And in a more accurate model, the dynamics are even {\it sub}-linear. In this {\muscle} prescription shown in Fig.\ \ref{fig:muscle_voids}, the behavior of a patch is entirely determined by a `stretch parameter' $\psi$ \cite{Neyrinck2013}, a displacement-divergence that behaves a bit like a density (which we denote, in units of the mean, by $\rho/\bar{\rho}$). In voids, where $\delta_{\rm lin}$ is the linear density,
\begin{eqnarray}
    \psi & = & 3\left[\sqrt{1-(2/3)\delta_{\rm lin}}-1\right],~{\rm derivable~from}\\
    \rho/\bar{\rho} & = &  \left[1-(2/3)\delta_{\rm lin}\right]^{-3/2}\label{eqn:bernard}
\end{eqnarray}
assuming an isotropic cell. Eq.\ (\ref{eqn:bernard}) is an accurate fit to the spherical expansion model in voids, found in Ref.\ \cite{Bernardeau1994}.

In voids, these variables depend on the initial density in the cell in a relationship {\it weaker} than linear (the derivatives with respect to $\delta_{\rm lin}$ of both expressions are less than 1 for $\delta_{\rm lin}<0$). Conceptually, this weak dependence occurs because in low-density regions, the expansion of the Universe is faster, and structure growth is suppressed; this is the opposite of chaos. It suggests even that if a density perturbation is introduced in a void at some non-initial epoch, that perturbation will not grow chaotically. We test this situation below.

In a neutralino-like dark-matter model, there is structure in the dark matter all the way down to scales below 1 comoving parsec, to $\sim$Earth-mass haloes \cite{DiemandEtal2005}. In this model, voids at the current epoch would have density minima reaching $\lesssim10^{-3}$ times the mean density \cite{YangEtal2015}. In such a place, the average distance from a dark-matter particle to a neighbor dark-matter particle could be $\sim 100$ meters, an amusingly human scale. Since predicability is highest at low density, perhaps the scale of predictability for individual dark-matter particles could be of that order.

\section{Cosmological simulations tracking chaotic dynamics}
\label{sec:dmosim}
We have given some reasons to think that the degree of chaos in single-stream regions is far suppressed compared to higher-density regions, but we quantify this explicitly here. Strictly speaking, chaos often means that a perturbation grows exponentially. But here we are looser with the definition, not trying to distinguish between exponential and e.g.\ power-law growth. Below, we compare the growths of various quantities, speaking loosely that if one rises faster than the other, it is `more chaotic.'

We follow the approach of \cite{GenelEtal2019}, who investigated the degree of chaos using cosmological $N$-body and hydrodynamic simulations. Like them, we use `shadow simulations' to quantify the degree of chaos, in which a simulation is duplicated at some timestep, machine-precision-level perturbations to particle positions are introduced in the duplicate, and distances between the same particles in different simulations are tracked as the simulations proceed. A particle's shadow is called that because it is a tracer in a different simulation that does not interact with it. We introduce position perturbations at redshift $z=5$, with random directions but amplitude 1 $\mu$pc/$h$ for all particles; this is of order the smallest perturbation possible in a double-precision position with maximum tens of Mpc.

This shadow distance can be thought of as a particle-by-particle estimate of the length scale of predictability $s_p(m,e, {\rm dark~matter},\allowbreak c_{N {\rm\mhyphen body}})$. This is because the perturbations introduced at $z=5$ are barely above machine precision, and so the perturbations might be thought of as a (bit overestimated) measure of the fundamental particle-position uncertainty in the simulation. 

The dark-matter-only (DMO) simulations use only the gravitational physics in {\scshape Arepo} \cite{WeinbergerEtal2020}, which for this purpose is essentially a run of the code {\scshape Gadget4} \cite{SpringelEtal2021}. The simulations assume a $\Lambda$CDM cosmological model and evolve a cubic volume of (25 Mpc/$h$)$^3$ over cosmic time (from cosmological initial conditions at $z=127$) using 256$^3$ particles.

This analysis was not as simple as we expected, because we found that the results depended sensitively on the gravitational force-softening prescription. Softening (reducing the gravitational force within a softening length) is necessary because simulations approximate the continuous (on astronomical scales) medium of dark matter with isotropic super-particles dozens of orders of magnitude more massive than actual particles. We note that this ambiguity and deficiency in standard $N$-body simulations can be remedied in principle in simulations that treat the matter as occupying a `dark-matter sheet' \cite{AbelEtal2012,ShandarinEtal2012,FalckEtal2012,HahnEtal2013,HahnAngulo2016,StuckerEtal2020} whose shape is traced with vertices at particle positions. In the future, we would like to repeat our analysis with such a code.

$N$-body simulations are usually calibrated for accuracy in haloes, dark-matter clumps where galaxies would form. We use a recently advocated prescription \cite{GarrisonEtal2021}, setting the softening to be $1/30$ of the initial interparticle spacing, and keeping it constant in physical rather than comoving coordinates.

\begin{figure}
\centering
    \includegraphics[width=0.49\columnwidth]{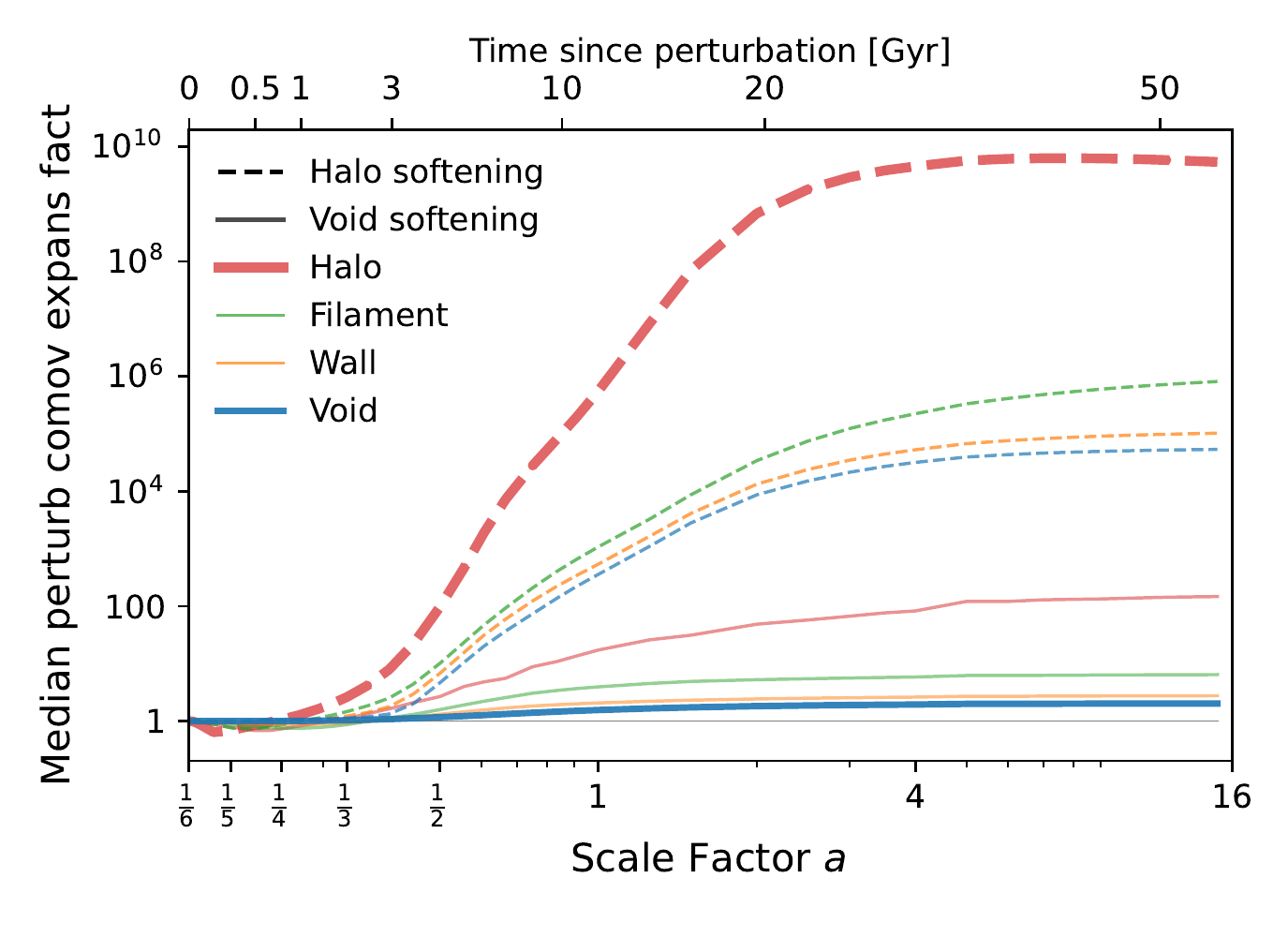}
    \includegraphics[width=0.49\columnwidth]{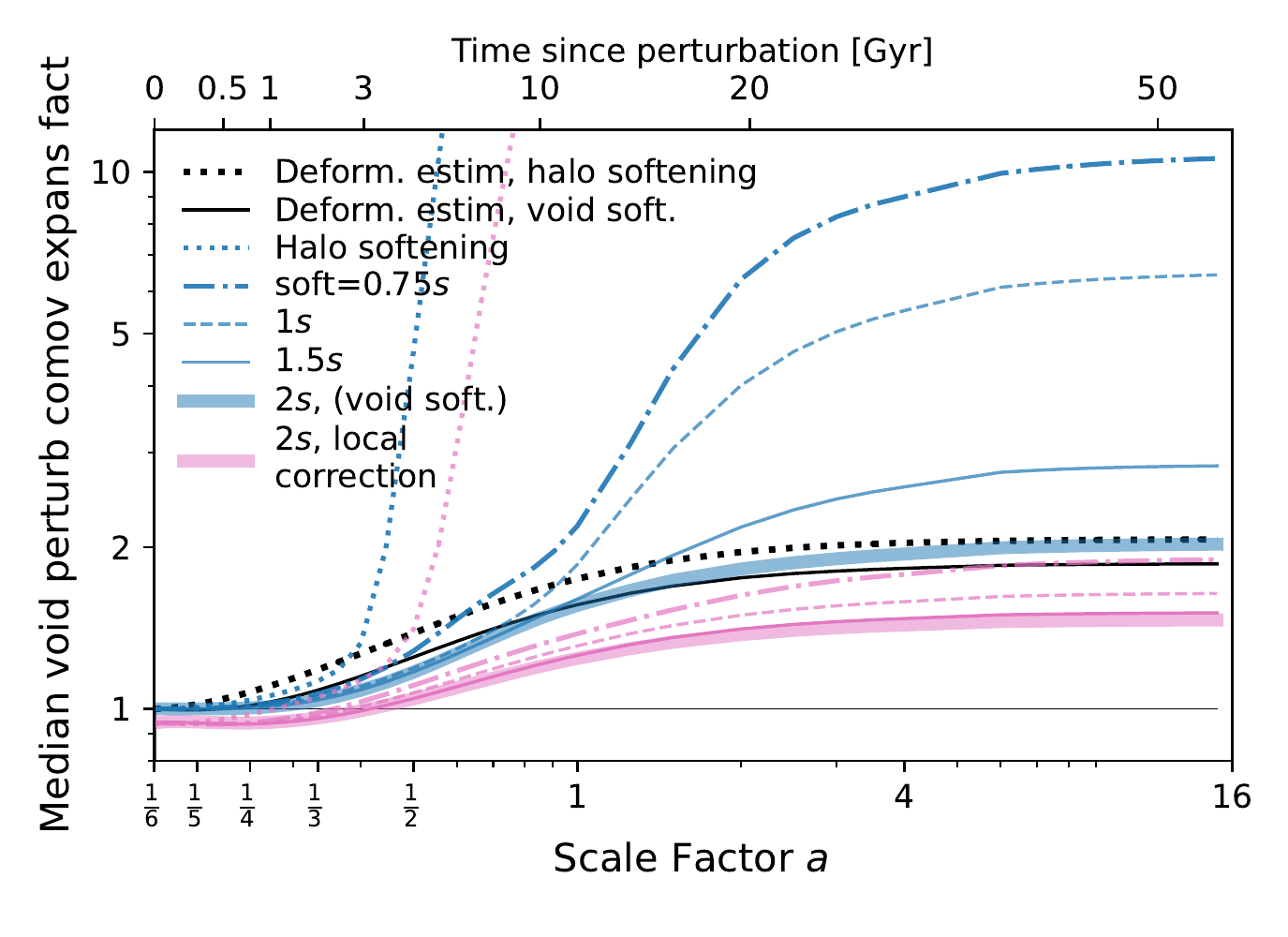}
    \caption{{\it Left}: Shadow separations in different cosmic-web morphology types. All particles are single-stream-void in the initial conditions, and collapses bring them higher in web type. Perturbations between pairs of shadow $N$-body dark-matter simulations start at $z=5$ ($a=1/6$), near machine precision (1 $\mu$pc/$h$) and expand typically to kpc-scale in haloes, at maximum even Mpc-scale. Curves show the evolution in time of the median expansion factor of the 1 $\mu$pc/$h$ perturbations, within each web type.\\
    Results differed substantially using a typical, small force softening tuned for haloes (`halo softening'), compared to a `void softening' tuned for void particle separations, of twice the initial lattice spacing. Naively using halo softening for halo particles, and void softening for void particles, at $z=0$ ($a=1$) there is a large factor of $\sim 10^5$ difference in median particle shadow-distance expansion between voids and haloes. At $a\sim5$, this grows to $10^9$.\\
    {\it Right}: The shadow distance as measured in voids using the void softening (solid blue) compared to estimates from the average distortions of void volume elements with much larger, void-range softening. The agreement of the pink curves at long softening, as well as the agreement in shape with the longest-softening blue curve, motivates the choice of void softening at twice the mean interparticle spacing.}

\label{fig:mediansofts}
\end{figure}

Using this halo softening prescription, as expected, we found that in haloes (as classified by \origami, \cite{FalckEtal2012}), particles' distance perturbations grow to many orders of magnitude greater than in voids (see left panel of Fig.\ \ref{fig:mediansofts}). In haloes, the perturbations grow to a median of many kpc, up to Mpc-scale, i.e.\ the sizes of the largest haloes. This particular curve shows power-law instead of exponential growth past $z\sim 0.5$, but we still call this chaos. One reason to do so is that we use the median over many haloes, of different sizes that limit the growth of their shadow distances. But we do note that the growth of some quantities in dark-matter dynamics physically may be power-law rather than exponential; Ref.\ \cite{Colombi2021} has found that the number of mesh elements required to resolve the full dark-matter sheet inside a halo experiences power-law and not exponential growth.

With `halo softening,' void shadow distances still grow by $\sim$5 orders of magnitude; should we call this chaos, too? We do not think so. To capture the dynamics accurately in voids, we need a much larger softening length, of order the typical particle separation there. If the softening length is much smaller than particle separations, the simulation proceeds as though the density is zero almost everywhere (between void particles). But with a softening length of order interparticle distances, the density estimate between particles is more sensible \cite[e.g.][]{StuckerEtal2020,Colombi2021}. 

We wish to test a hypothesis that the growth in shadow distances in voids is driven simply by deformations in the particle lattice, not by chaos. In a void, a cube of particles outlining a volume typically expands (even in comoving coordinates, since it is growing underdense), driving shadow particles with tiny separations slightly apart. We use this idea to predict the growth of shadow distances, first by measuring the distortion tensor at each particle $D_{xq}$. We measure this with finite differences of the displacement field in Lagrangian coordinates \cite{StuckerEtal2020}. A small initial displacement $\Delta \vec{q}$ will then result in a final displacement $\Delta \vec{x} = D_{xq} \Delta \vec{q}$. Using this, we measure the median growth of random displacements with random directions by Monte-Carlo sampling $|\Delta \vec{x}| / |\Delta \vec{q}|$ in the simulation.

The right panel of Fig.\ \ref{fig:mediansofts} shows that with a sufficiently large smoothing length of twice the mean spacing $s$, the median perturbation expansion factor resembles this deformation estimate. A further `local correction' brings these curves (in pink) nearly together, causing them to converge to roughly the shape of the deformation estimate for smoothing lengths $\gtrsim 1.5s$.\footnote{This correction subtracted off possible large-scale displacements from noise. We averaged the three components of the shadow separation in Lagrangian coordinates of particle neighbors in a Gaussian kernel with $\sigma$ set to one particle spacing, and subtracted this average flow from the shadow separation of each particle.} This `local correction' evidently overcorrects a bit, dipping below unity initially. Still, this analysis strongly suggests that the growth of position perturbations in voids is not a chaotic process, but is more related to deformations of the particle lattice. The deviations from that prediction are small, considering that we are numerically tracking $\mu$pc-scale separations in a 25 Mpc$/h$-large cosmological box.

Note that in Fig.\ \ref{fig:mediansofts}, we have traced the perturbation expansion into the rather far future, to $a=16$. The various curves saturate at $a\sim 4$. This happens because in $\Lambda$CDM, the universe transitions to exponential expansion around then, and the cosmic web freezes \cite{BushaEtal2003}. Interestingly, in the left panel, the top, halo perturbation curve {\it decreases} with time past $a\sim 6$. A clue for why this happens is the spatial scale of this extreme expansion factor; the median has reached the scale of several kpc, with maximum even a few Mpc. As cosmic structure freezes out, haloes contract in comoving coordinates. The downturn here suggests that the perturbation expansion factor of a particle is limited mainly by the physical size of its halo. When structure freezes, haloes cease to merge, and contract, pulling the dashed red curve downward. The filament perturbation expansion factor is the only curve that continues to ascend appreciably at high $a$; perhaps that is because particles, despite being confined in a narrowing filament, are still a bit free to flow and separate along the filament.

There is a caveat that these results doubtless carry some resolution dependence. In a given void, higher resolution or initial power typically produce additional filaments and walls (density ridges) \cite{AragonCalvoSzalay2013,YangEtal2015}. But deep in a void, these often do not manage to collapse and multistream, which is our key criterion for chaos arising. Even with full initial power in a CDM model, substantial volumes of the universe remain single-stream \cite{AnguloWhite2010,StuckerEtal2018}. So this resolution dependence is unlikely to change our conclusions qualitatively.

In voids, the scale of predictability is short (i.e., high accuracy is possible) even with little computation, and this is also where chaos seems never to occur. But the scale of predictability is large within the cosmic web, and persists at a large (albeit smaller) value even with high amounts of computation (that of an $N$-body simulation).

In filaments and walls, the curves are between those in voids and haloes, as expected. Interestingly, it is much closer to voids than haloes, its perturbations growing only a bit more than in voids. However, the distributions are roughly lognormal and overlapping. The distribution is especially wide for filament particles; some of these grow nearly as fast as for haloes. It would be interesting to test if a particle's degree of chaos might correlate more strongly with its number of `flip-flops,' or parity inversions of its local volume \cite{ShandarinMedvedev2017}, than cosmic-web type.


\section{Chaos and randomness in galaxies}
\label{sec:galaxies}
In the $N$-body simulations discussed in the last section, we found that the degree of chaos in dark-matter haloes was high even using only gravitational physics, finding that eventually, a $\mu$pc-scale perturbation at $z=5$ causes extreme uncertainties in dark-matter-particle positions, up to the largest halo diameters. The purely gravitational Hamiltonian dynamics of a collisionless dark-matter universe should conserve information perfectly in principle, but chaos in particle positions is known to be generally quite high in haloes \cite[e.g.][]{Suto1991,ThiebautEtal2008,KellerEtal2019,GenelEtal2019}. But dark-matter orbits still appear to sample the overall halo structure accurately, even if particles' positions in their orbits can be inaccurate \cite[e.g.][]{StuckerEtal2020}.

The amount of randomness explodes when hydrodynamics and numerically random star-formation and feedback processes are included. To investigate this, we ran the same initial conditions with the star formation and feedback model calibrated for the state-of-the-art IllustrisTNG simulations \cite{PillepichEtal2018}, using the \arepo\ code; we abbreviate these as `TNG'. The main physics processes included in this model are radiative gas cooling, star formation out of dense gas, injection of energy and mass (including heavy elements) into the surrounding gas by evolving stellar populations, as well as formation and growth of supermassive black holes, and their injection of energy into the surrounding gas according to their accretion rate.

\begin{figure}
\centering
    \includegraphics[width=\columnwidth]{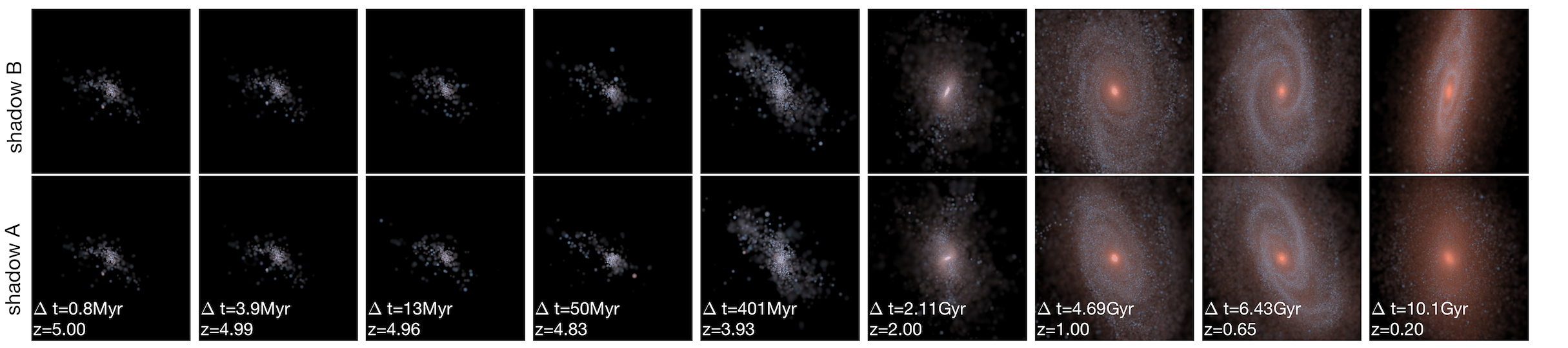}
    \caption{Machine-precision shadow displacements such as we use at redshift $z=5$ using the TNG hydrodynamics, star-formation, and outflow models cause qualitative morphological changes in some galaxies by redshift $z=0$. Taken from Fig.\ 1 of \cite{GenelEtal2019}, used with permission. See also \url{https://youtu.be/ftjqPnbLJ28}.}
    \label{fig:shadow_galaxies}
\end{figure}

Because of the finite dynamic range of resolution, all cosmological simulations that include star-formation, feedback, and outflow prescriptions do so with sub-grid modelling, whose details depend on computer-generated (pseudo)random numbers. Steps are taken in the simulation to make each run in principle repeatable and deterministic, using e.g.\ the same sequence of random numbers, but in the simulation, the details of each outflow are generated at random, not specified in the initial conditions. The level of stochasticity is dramatic even on galaxy scales, contributing to a fraction of quite different-looking galaxies at the final epoch, as shown in Fig.\ \ref{fig:shadow_galaxies}.

How much of this randomness exists in the real Universe? Most likely, the simulations have much more stochasticity than reality. In the TNG model, when star formation happens, it does so in the form of giant $\sim 10^7 M_\odot$ superparticles, introduced all at once, with a somewhat stochastic velocity, rather than in a more smoothly generated distribution of many stars, generated over Myrs or Gyrs, as would happen in reality. On the other hand, there are some aspects of the simulation that may suppress stochasticity; an example is AGN accretion, which is implemented in TNG smoothly and without random numbers: in every timestep, the black hole absorbs a set fraction of the mass in its hydrodynamic cell.

But physical stochasticity may be surprisingly common, as well.
A prime example of a process that might broadcast perturbations to galactic scales and beyond is a turbulent jet from an active galactic nucleus (AGN).

Suppose that there are random perturbations in a turbulent accretion disk. The event horizon is a bottleneck in scale in the accretion/jet process; if non-primordial randomness has arrived there during accretion, it could eventually reach large scale or amplitude in the jet. These perturbations could reach far out into the galaxy, and even outside it, since outflows are observed to reach, in their extremes, Mpc-scale.

Even before an interaction with a black hole that launches a jet, accretion disks are subject to numerous instabilities; some observed variability in AGN or X-ray binaries is well-modeled by random viscosity fluctuations, physically arising from a process like turbulence \cite{TurnerReynolds2021}. The emission from a prototypical accreting stellar-mass black hole, Cygnus X-1, varies on timescales as short of 16-100 ms \cite{SkipperEtal2013}, and up to years or decades \cite{KaritskayaEtal2019}.

Another example of a (mostly, except for mention of the Planck length) classical system that can be `fundamentally unpredictable' and non-reversible is in the three-body problem. It is well-known to have chaotic regimes, but Ref.\ \cite{Boekholt2020} found that in $\sim 5$\% of idealized, zero angular momentum, triples of astrophysical black holes, the system is so chaotic that Planck-length differences in ICs eventually get amplified to astronomical scales in black-hole positions. If these processes have some non-primordially random component, all this randomness being continuously sprayed into the galaxy could add up to be substantial.

It is difficult to assess the magnitude of non-primordial randomness that might be introduced in such systems. Interestingly, a technique called recurrence analysis \cite{EckmannEtal1987} is capable of empirically distinguishing some types of true randomness from simple versions of deterministic chaos. It works by detecting recurrences (if not strict periodicity) of patterns in data such as time series, e.g.\ chaotic strange attractors. These tools are starting to be applied in astronomy, e.g.\ to stellar orbits \cite{SchaffnerDaniel2022}. Applied to AGN, some systems have been found to be well-described by deterministic chaos, and others better-described by stochastic models such as a damped random walk and a stochastic damped harmonic oscillator \cite{SukovaEtal2016,MorenoEtal2019,PhillipsonEtal2020,YuEtal2022}. Admittedly, finding that a set of data is best-described as stochastic says little about the source of that stochasticity; it could conceivably just be linked to initial information in a way too complex to pick up with this method. And it is interesting that many systems are best-described by deterministic chaos, when in principle everything could be quite random. Still, the population of stochastic systems is an interestingly empirical sign that in many AGN, randomness (or, at least, numerical `stochasticity,' in a more limited sense) exists.

\subsection{Spontaneous stochasticity}
\label{sec:ss}
We have been discussing non-primordial randomness as arising from quantum processes after inflation. But there is another possible type of {\it classical} non-primordial randomness that could appear in galaxies, arising from arbitrarily small perturbations in sufficiently turbulent flows, called `spontaneous stochasticity,' also called the `real butterfly effect' \cite{PalmerEtal2014}. This may amplify even the tiniest bits of introduced randomness to galactic scales, more than is commonly appreciated.

Spontaneous stochasticity was essentially discovered by Richardson \cite{Richardson1926}, although the concept has evolved, and it acquired that name rather recently \cite{FalkovichEtal2001}. Ref.\ \cite{Kupiainen2003} describes Richardson's simple, human-scale experiment, which happened to occur around when the spookiness of quantum mechanics was being unveiled. Richardson released pairs of balloons in the turbulent atmosphere with different initial separations. He found that their separations $r(t)$ as a function of time behaved as $r(t)^2\sim A(r_0)t^3$, where $A(r_0)$ is a constant depending on initial separation $r_0$. The remarkable thing was that for sufficiently small $r_0$, $A(r_0)$ was independent of $r_0$, in particular nonzero as $r_0\to 0$. This implies that even if one could release two balloons from the exact same location, the randomness from turbulence would still typically make them diverge from each other! There is therefore a range of possible final conditions from the same initial conditions, i.e., arguably, non-determinism.

A physical and philosophical question is whether spontaneous stochasticity comprises randomness as `true' as quantum randomness. Does `God play dice' to produce the trajectories of particles in a highly turbulent flow? Physically, if a turbulent flow may take multiple paths, it still only takes one, and there is no indication of the other possibilities. `Dice rolls' are undetectable. But there is evidence of Richardson's law and spontaneous stochasticity in various turbulent systems, analytically \cite{Eyink2011} and numerically \cite{SawfordEtal2008,Eyink2010,LazarianEtal2015,BiferaleEtal2018}. For example, the recent Ref.\ \cite{ThalabardEtal2020} found evidence for spontaneous stochasticity in simulations of the the well-known Kelvin-Helmholtz instability; over a few ranges of perturbation strengths, the sizes of energy perturbations reached the same asymptotic large values, regardless of the size of initial perturbations introduced. Spontaneous stochasticity may allow high-energy phenomena such as magnetic reconnection in astrophysical plasmas \cite{EyinkEtal2013}. Theorems implying determinism and a unique solution in classical fluid dynamics do not apply to a sufficiently turbulent flow, if the velocity field is `rough.'\footnote{`Rough' means continuous only in a H\"older sense: there exists constants $C>0$ and $0<h<1$ such that $|v(x_1)-v(x_2)|\le C|x_1-x_2|^h$ for the velocity field $v$, for all positions $x_1$ and $x_2$. Additionally, this condition does not hold for $h\ge 1$.} Some astrophysical turbulent velocity fields are known to be rough in this way \cite{EyinkEtal2013}.

Even if classical turbulent fluid dynamics does not produce genuine randomness in itself through spontaneous stochasticity, it is worth asking whether this mechanism of eventually amplifying arbitrarily small fluctuations to large size might allow tiny quantum fluctuations to expand to become large enough to ultimately influence where stars form in a galaxy, more than is usually appreciated.  Ref.\ \cite{EyinkDrivas2015} suggests that quantum randomness propagating to larger scales could physically source spontaneous stochasticity.

\section{Can galactic outflows pollute voids with chaos or randomness?}
\label{sec:outflows}
Consider a dark-matter-only universe. Single-stream regions remain pristine, unvisited by matter ejected after collapse from galaxies. This is in fact a tautology, since such an ejection would constitute multistreaming. But more meaningfully, the multistreaming, or collapsed, structures observed in simulations tend to be quite compact compared to the scales in between structures such as dark-matter haloes. One way to understand this is after infall, the energy a particle has acquired during infall might carry it out as far from the structure as it started in physical units, but not as far in comoving units, because of the universe's expansion.

One exception to this picture would be if the dark matter has an astronomically appreciable mass, e.g. if it were comprised of black holes. Then, they could chaotically scatter and (perhaps rarely) escape from collapsed structures with much higher velocity than typical infall velocities, and voyage intergalactically. In this case, the scattering events could broadcast chaotic (or random, \cite{Boekholt2020}) information from within galaxies deeply into single-stream regions.  But for the purposes of this paper, we will assume a type of dark matter that would not have such DMO scattering events (such as WIMPs or axions).

But more importantly, we know that baryonic outflows such as AGN jets routinely eject matter outside of galaxies. Might these perturbations incite chaos in voids?

We already suspect from \S \ref{sec:dmosim} that small perturbations generally do not grow chaotically for dark-matter particles. But it is still interesting to see the effect of possibly random outflows on the determinism in voids explicitly; here, we examine results from the {\scshape Arepo} TNG model described in \S \ref{sec:galaxies}.

Although we have been discussing dark matter, it is probably more interesting to consider the effect of baryonic outflows on baryons in voids. Unfortunately, though, the baryonic and star particles in the simulation are not truly mass-tracking Lagrangian tracers as the dark-matter particles are, because the hydrodynamic model uses cells which exchange mass. It is possible to construct Lagrangian tracers for baryons \cite{GenelEtal2013}, but this method is itself stochastic, and would not test for chaos. In this first exploration of determinism in voids, we concentrate on the practically more straightforward question of how perturbations might grow in dark matter in voids.

\begin{figure}
\centering
    \includegraphics[width=\columnwidth]{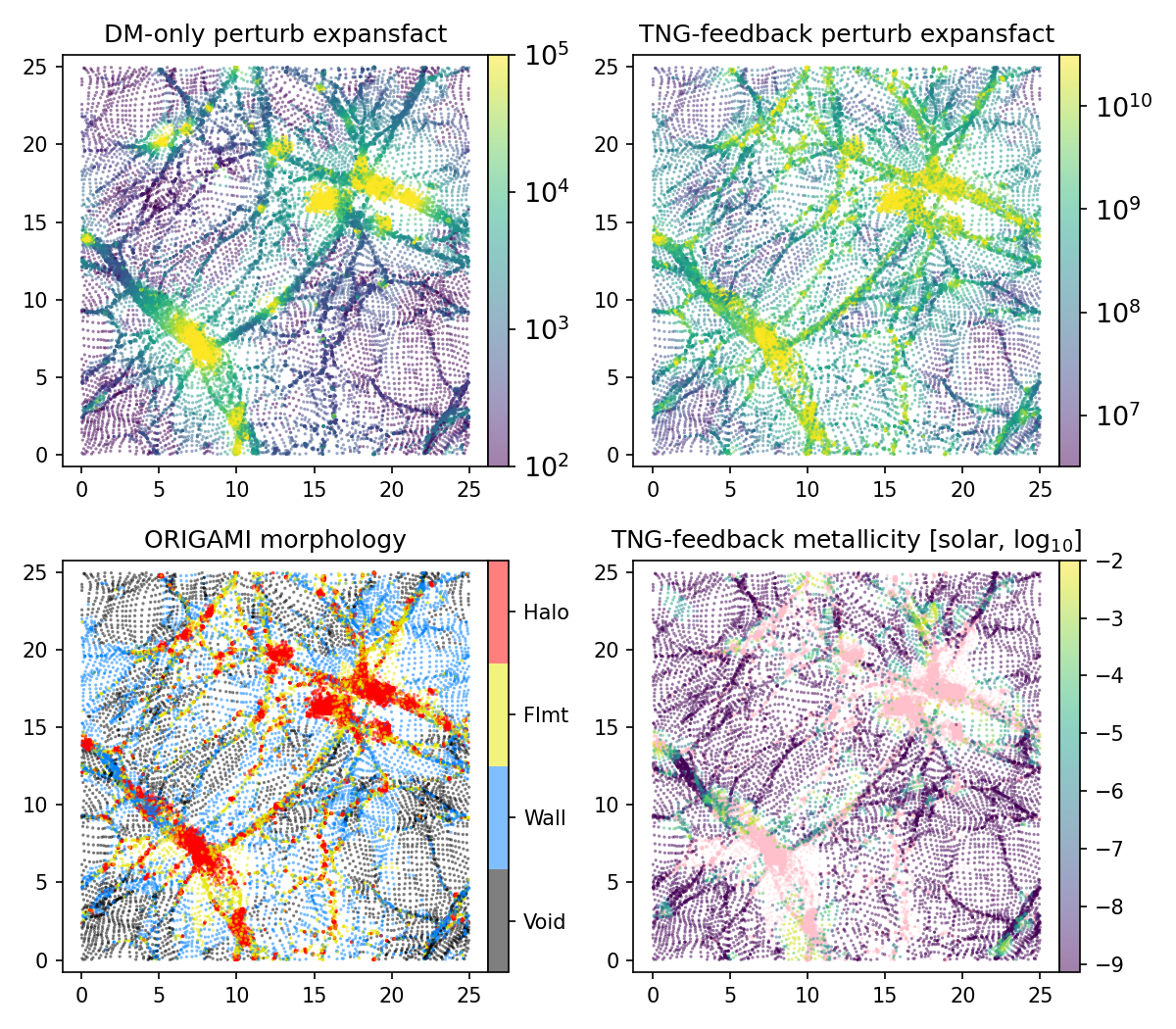}
    \caption{Visualizations of levels of chaos (factors by which machine-precision-level perturbations grow from redshift $z=5$ to $z=0$) for a $256^2$ sheet of dark-matter particles in their cosmic-web contexts, and the effect of hydrodynamics and star-formation feedback.\\
    {\it Upper left}: Perturbation expansion factor $f$ in a dark-matter-only (DMO) simulation. Note that colorscales here and in the next panel are saturated at the high end, but a few more orders of magnitude are shown in Fig.\ \ref{fig:shadows_metals_lag}, where halo particles are distinguishable.\\
    {\it Upper right}: $f$ in a simulation with the same initial conditions, but that uses a TNG model of hydrodynamics and star-formation feedback. Note the factor of $\sim 10^5$ difference in the color scale.\\
    {\it Lower right}: Dark-matter particle metallicities (of the nearest gas particle) for wall and void particles in a simulation with a TNG model incorporating hydrodynamics and star-formation feedback. Pink particles have filament or halo \origami\ web types.\\    
    {\it Lower left}: Particles' \origami\ web-type classification, from the DMO simulation.\\  
}
\label{fig:shadows_metals_eul}
\end{figure}

Fig.\ \ref{fig:shadows_metals_eul} shows a 2D slice of $256^2$ dark-matter particles from the simulation at $z=0$, colored in various ways, including the `perturbation expansion factor' at top, which is the ratio between particle shadow-pair distances at $z=0$ and $z=5$ (this is the same quantity as in the $y$-axis of Fig.\ \ref{fig:mediansofts}). We use the metallicity at the nearest baryonic particle (equivalently, of the hydrodynamic cell that the dark-matter particle inhabits), in the bottom-right, as a measure of whether a baryonic outflow has passed a particle, and how strong that outflow was. We show metallicities only for void and wall particles. We did not estimate metallicities for filament and halo particles because of the high degree of multistreaming, making the connection between each dark-matter particle and the hydrodynamic cell it occupies possibly tenuous. 

\begin{figure}
\centering
    \includegraphics[width=\columnwidth]{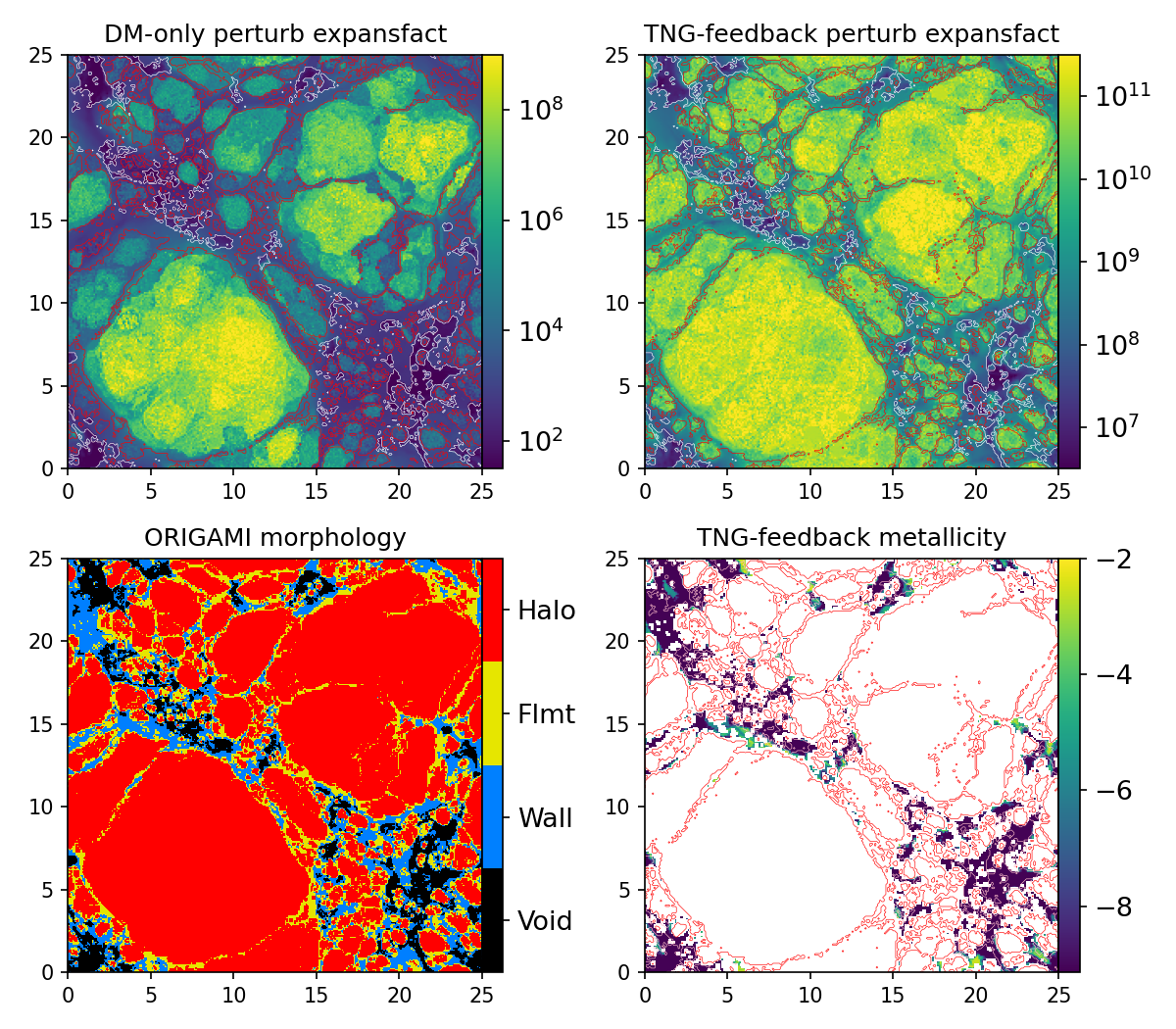}
    \caption{A Lagrangian view of the $256^2$ sheet of particles shown in Fig.~\ref{fig:shadows_metals_eul}, with one pixel per particle, shown in its initial position on the cubic lattice of particles. Note the difference in color scale in the upper two panels. In both, the upper end is nearly saturated, but there is still another order of magnitude cropped off at the upper end, for clarity. Faint white contours indicate boundaries between void and wall regions, and thin, faint red contours indicate boundaries between filament and halo regions. At bottom right, halo and filament particles/pixels are simply colored white.}
\label{fig:shadows_metals_lag}
\end{figure}

In Fig.\ \ref{fig:shadows_metals_lag}, we show the same view in Lagrangian space, i.e.\ where each particle is shown as a pixel in its location on the initial 256$^3$ grid. Here, haloes in Fig.\ \ref{fig:shadows_metals_eul} have expanded to the comoving patches that collapsed to form them. This view reveals substructure in haloes; the central, likely first-collapsing parts have the greatest DMO separations between shadow simulations. In the TNG case, though, haloes show more uniformly high separations. Curiously some of the highest patches in DMO are relatively low in TNG. Here and in the Eulerian view, we applied the `local correction' only for void particles, since it takes into account only Lagrangian neighbors. With the corrections, the perturbation expansion factors are deeply lower than in surrounding walls (with boundaries indicated by faint white contours in upper panels). Without the correction, these regions are just a bit less suppressed, but we do not show a comparison for simplicity.

The perturbation expansion factor for the dark-matter particles is typically, and rather globally, 3-5 orders of magnitude larger in the TNG simulation than in the DMO simulation\footnote{There is only a bit smaller difference in haloes because some of the dark-matter particles in the DMO case are already near their halo diameter limit of $0.1$-$1$ Mpc/$h$, which give a separation expansion factor of $10^{11-12}$.}. To try to track down the source of this large difference, we also ran simulations using two other sub-grid physics prescriptions; one `SP' (simple physics), with standard cooling and star formation, but no feedback (no galactic winds, black holes, or AGN) and one `NR' (no radiative cooling), with `adiabatic' hydrodynamics, but no cooling or star formation.

The particles had high shadow separations in the SP case, only a bit less than in TNG, so it is not the feedback that excites the rather global shadow separations. The NR simulation had shadow separations more noticeably suppressed compared to TNG, but was much closer to TNG than DMO. Importantly, the NR simulations did not have any explicit random processes. So, it seems that the numerical hydrodynamic prescription itself, rather than randomness explicitly injected into the simulation, largely produces the large shadow separations with hydrodynamics. This is unfortunate, but these simulations still may provide state-of-the-art estimates of chaos in different regimes. Particularly, we do not have a good reason to doubt the physical meaning of relative TNG shadow-separation magnitudes for different particles.

\begin{figure}
\centering
    \includegraphics[width=0.6\columnwidth]{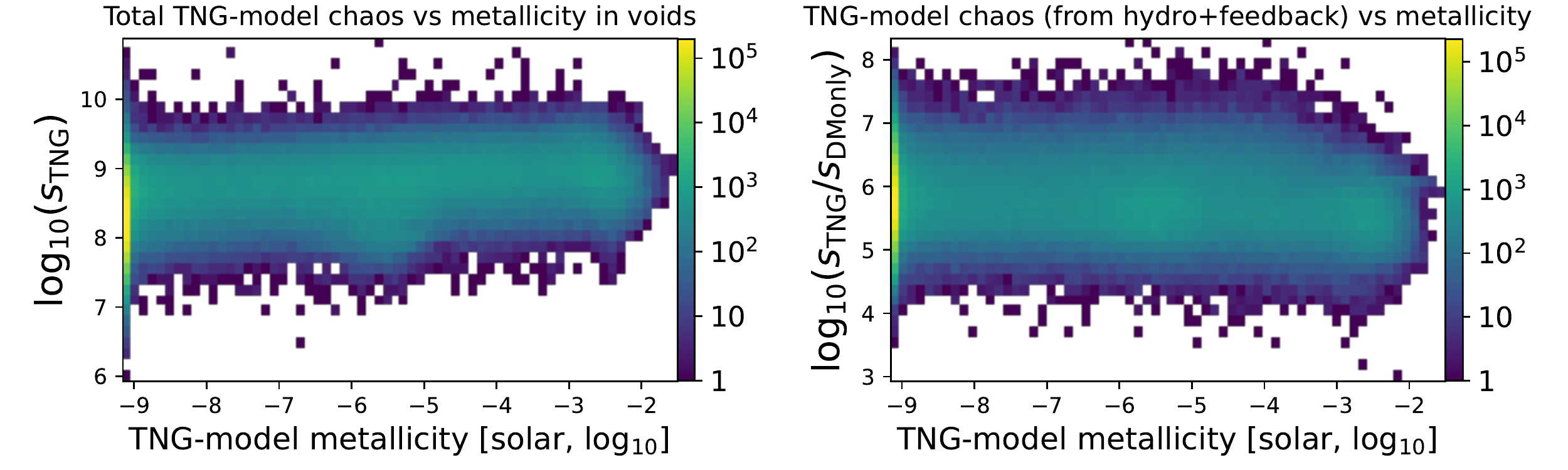}
    \caption{2D histograms comparing the level of `chaos' (the shadow-separation expansion factor $s$) in voids, for both low or `zero' ($10^{-9}$)-metallicity regions, and for higher-metallicity regions of voids, presumably most affected by outflows.\\
    {\it Left}: The TNG-model shadow-separation expansion factor as a function of metallicity.\\
    {\it Right}: The same, except attempting to remove effects of the most distorted single-stream cells lying near collapsed structures. We simply divide the quantity in the left panel by the measured DMO expansion factor.}
\label{fig:chaos_vs_metallicity.png}
\end{figure}

Visually, there seems to be little connection between the level of chaos (measured by the perturbation expansion factor) and metallicity for particles in voids. Indeed, this shows up in Fig.\ \ref{fig:chaos_vs_metallicity.png}, in 2D histograms of expansion factors against metallicity. The raw TNG expansion factor shows a very gradual upturn with metallicity, but this is plausibly because the cells even in the DMO situation grow slightly more anisotropic near collapsed structures. After dividing out the DMO perturbation expansion factor, there is even a gradual downturn with metallicity. This is a crude correction; one problem is that the DMO and TNG expansion factors were computed with `void' and `halo' smoothing, respectively, making them possibly hard to compare meaningfully. Also, the operation of simply dividing the two distances glosses over differences in directions and full trajectories. But still, we suspect that the histogram at right is more meaningful than that at left, and we have no clear evidence that dark-matter regions of high metallicity are more chaotic.

We would need more in-depth analysis to understand if TNG outflows ever incite chaos in dark matter; our analysis does not rule out that this may happen in some cases. These may also be underrepresented because of the small box size of our simulations. But according to this preliminary work, and the previous measurements of low chaos in the DMO simulations, it seems that chaos is rare in the dark matter in voids, even where they have been polluted by outflows.

\section{Practical or physical relevance of the cosmic information budget}
\label{sec:dynamicalinfo}
We have been considering whether non-primordial randomness may contribute `information' to the cosmos. Conceptually, this information is whatever data the arrangement of matter at some epoch depends on. This information budget has philosophical interest, we believe, but also, as we discuss in this section, some relevance for doing practical science, and, speculatively, even to dynamics.

Kolmogorov information is the shortest set of data (and program code, which we neglect) needed to predict the final structure to some high accuracy. This data includes the ICs, as well as non-primordial randomness from information sources. Possibly, information sinks exist as well, if the total information may be rearranged to give the same structure \cite{Neyrinck2015}. This is a sort of entropy, since many microstates may give the same observed macrostate. But this leaves quite ambiguous what precisely is `observed' and on what length scales, and also why `observation' should be so crucial to this quantity.

This information budget has some practical scientific relevance, for `field-level' approaches to inferring cosmological parameters from observed galaxy surveys or other cosmological observables \cite[e.g.][]{SchmittfullEtal2019,AndrewsEtal2022}, i.e.\ simulating the full large-scale structure field including cosmological parameters, paying attention only to parameters of interest. It is also relevant for doing science with summary statistics; many commonly used statistics fail to capture the large-scale structure well on small scales \cite[e.g.][]{rh05,CarronNeyrinck2012}; with added randomness, summary statistics may lose sensitivity even in principle to physics being investigated.

The cosmic information budget is also relevant to holographic theories \cite[e.g.][]{BoussoEtal2007,EganLineweaver2010}. Ref.\ \cite{CortesEtal2022} argues that the entropy (or information) from life has come to dominate the cosmic information budget. If so, most of the current information would be non-primordially random; even more so, if turbulence and feedback processes generate some randomness, as discussed here.

There have even been some (admittedly `outrageous' \cite{CortesEtal2022}) suggestions that information may play a cosmological role, even as an explanation for the coincidence problem that dark energy turns on roughly when stars form appreciably, through a change in an entropy budget that could affect the expansion rate \cite[e.g.][]{EassonEtal2011,Gough2011,CapozzielloLuongo2018,GarciaBellido2021} This scenario is quite speculative; we find it implausible that small-scale changes in matter configurations could have global consequences. But quantifying the information budget is necessary to test this scenario, and processes such as we investigate here could contribute significantly. Perhaps, there could even be observable variations in the local expansion rate that depend on the local star-formation, or information-production rate.

\section{Summary and Discussion}
\label{sec:summary}
In this paper, we ask the question of where dynamics in the Universe is chaotic, and, furthermore, speculatively, where it is even non-deterministic. The first question, of where there is chaos, is straightforward to answer. In single-stream voids, where a dynamical time does not exist because matter has not stream-crossed or shell-crossed, the dynamics are generally predictable to high accuracy. This is definitely true for the dark matter, and possibly for the baryons as well. It is ironic that it is only in the voids, where it is almost impossible to observe the density field at high resolution in the real Universe, that initial information is pristine.

Advancing through the cosmic web types of walls and filaments, the degree of predictability in individual particle positions decreases (the scale on which one can predict them increases). Haloes form a boundary of practical predictability, since a microscopic perturbation to the whole ensemble of particles at early times can lead to kiloparsec or megaparsec-scale differences in their positions at late times. We refer to the behavior in haloes loosely as `chaos', even though the amount of predictability and possible exponential growth of perturbations to the system depends on aspects such as the softening length, and algorithm. With more dependence on the definition, patches of filaments and walls, as well, may be chaotic. We must note, though, that in the dark matter, the structures `look' the same, that is, it appears that particles still sample the phase-space density rather well, even if individual positions are chaotic. The same may not be true for the gas and baryons, though. 

It would be interesting to test the predictive capacity of Vlasov-Poisson simulations that model the dark matter as a deforming sheet. We suspect that these would show less chaos, since they would not be subject to discrete particle scattering, and also because they would not be subject to the ambiguity of a global softening length that we found to affect the growth of position perturbations quite sensitively.



In a universe with only collisionless dark matter, there is no reason to doubt that the ultimate `scale of predictability' is zero, i.e.\ that such a universe manages to compute its structure without ambiguity or randomness introduced after the primordial fluctuations are imprinted. But it is worth asking, as we do in this paper, whether the dissipational and possibly random physics of baryons might produce a nonzero, or even large, asymptotic value of the uncertainty in baryonic or even (through e.g. dynamical friction) in the dark matter, in galaxies. We do not know the answer to this, but because of high-energy phenomena such as AGN that can propagate small-scale fluctuations up to galactic scale, we do not think that this question is trivially yes or no. Some further work that might help to quantify the magnitude of the fundamental uncertainty could be to examine in detail the chaos and possible sources of true stochasticity in astrophysical accretion disks and jets. But even if non-primordial randomness is negligible on astronomical scales, it is unlikely to be precisely zero. Randomness abounds on Earth; by the Copernican principle, our planet is likely not the only place where it abounds.

We thank Greg Eyink, Marina Cort\^es, Stu Kauffman and Greg Bryan for illuminating and encouraging discussions. M.N.\ acknowledges support by the Spanish grant PID2020-114035GB-100 (MINECO/AEI/FEDER, UE), and Aspen travel support by the Simons Foundation. S.G.\ also acknowledges support by the Simons Foundation, at the Flatiron Institute.
J.S. acknowledges funding from the European Research Council (ERC) under the European Union’s Horizon 2020 research and innovation program with grant agreement No. 716151 (BACCO). This work was finished at the Aspen Center for Physics, which is supported by National Science Foundation grant PHY-1607611. 

\footnotesize
\bibliographystyle{spphys_incltitle.bst}
\bibliography{refs.bib}

\end{document}